\documentclass[a4paper,10pt]{article}
\usepackage[utf8x]{inputenc}
\usepackage{graphicx}
\usepackage{subfigure}
\usepackage{amssymb}
\title{\bf An intermediate framework between WIMP, FIMP, and EWIP dark matter}
\author{Carlos E. Yaguna\footnote{carlos.yaguna@uni-muenster.de} \\ 
\small \it Institut f\"ur Theoretische Physik, Universit\"at M\"unster,\\
\small  \it Wilhelm-Klemm-Stra\ss e 9, D-48149 M\"unster, Germany}
\date{}
\newcommand{\mgrav}{m_{\tilde G}}
\newcommand{\mglu}{m_{\tilde g}}
\newcommand{\trh}{T_{RH}}

\newcommand{\oh}{\Omega h^2}
\newcommand{\gev}{\mathrm{~GeV}}
\newcommand{\tev}{\mathrm{~TeV}}
\newcommand{\sv}{\langle\sigma v\rangle}
\newcommand{\svunits}{\mathrm{cm}^3\mathrm{s}^{-1}}

\begin{document}
\maketitle

\begin{abstract}
WIMP (Weakly Interacting Massive Particle), FIMP (Feebly interacting Massive Particle) and EWIP (Extremely Weakly Interacting Particle) dark matter are different theoretical frameworks that have been postulated to explain the dark matter. In this paper we examine an intermediate scenario that combines features from these three frameworks. It consists of a weakly interacting  particle --\'a la WIMP-- that does not reach thermal equilibrium in the early Universe --\'a la FIMP-- and whose relic density is determined by the reheating temperature of the Universe --\'a la EWIP. As an example, an explicit realization of this framework, based on the singlet scalar model of dark matter, is analyzed in detail. In particular, the relic density is studied as a function of the parameters of the model, and the new viable region within this intermediate scenario is determined.  Finally, it is shown that this alternative framework of dark matter allows for  arbitrarily heavy dark matter particles and that it suggests a connection between dark matter and inflation. 
\end{abstract}

\section{Motivation}
The nature of  dark matter is one of the most important open problems in fundamental physics today. We know that dark matter exists, for the evidence in its favor is overwhelming, but we still ignore the identity and the properties of the dark matter particle. Since none of the known particles can play the role of the dark matter, the solution to this puzzle  requires new physics, physics beyond the Standard Model.

WIMP (Weakly Interacting Massive Particle) dark matter is a generic framework that can naturally explain the observed dark matter density. Its basic idea is that a stable massive particle ($M\sim 100-1000$ GeV) with weak-strength interactions will typically have, via freeze-out in the early Universe, a relic density not far from the observed dark matter density. Given that several extensions of the Standard Model  contain such a particle, they can account in a natural way for the dark matter of the Universe. In fact, most of the dark matter models considered in the literature, including the neutralino \cite{Griest:2000kj}, the lightest KK particle \cite{Hooper:2007qk} and scalar dark matter \cite{Honorez:2010re, Yaguna:2008hd},  fall within this category. And most of the experimental effort in dark matter detection is focused on this kind of candidates \cite{Bertone:2004pz}. It must be stressed, however, that currently there are no indications that  dark matter is actually composed of WIMPs. It is important therefore to consider  viable alternatives to this paradigm.

One interesting alternative is FIMP (Feebly Interacting Massive Particle) dark matter \cite{Hall:2009bx,Yaguna:2011qn}. In this case the interactions of the dark matter particles are so suppressed that they are unable to reach thermal equilibrium in the early Universe. They are slowly produced as the Universe cools down but are never abundant enough to annihilate among themselves. To reproduce the observed value of  the  dark matter  density, the coupling between the dark matter and the thermal plasma should be of order $10^{-11}$-$10^{-12}$. As a result of such feeble interactions, FIMPs are not expected to produce significant signals at direct or indirect detection experiments. It is, nonetheless, a framework as simple and predictive as the WIMP one.  

Another alternative is  EWIP (Extremely Weakly Interacting Particle) dark matter \cite{Choi:2005vq, Steffen:2008qp}, sometimes called EWIMP. It features dark matter particles with extremely weakly and non-renormalizable interactions. The gravitino and the axino are two well-known examples belonging to this class. Due to their non-renormalizable interactions, the dark matter  production is dominated by the high temperature regime and the relic density is ultimately determined by the reheating temperature of the Universe. 

In this paper, we examine a further alternative to the WIMP paradigm, an intermediate framework that share features from WIMP, FIMP and EWIP dark matter.  The scenario consists of a weakly interacting particle --\'a la WIMP-- that does not reach thermal equilibrium in the early Universe --\'a la FIMP. To achieve this, a reheating temperature smaller than the mass of the dark matter particle, though not necessarily \emph{low} in the conventional sense, is postulated. In consequence, the dark matter relic density turns out to depend also on the reheating temperature of the Universe --\'a la EWIP. We not only describe this intermediate framework but also analyze in detail one explicit realization of it based on the singlet scalar model of dark matter. In particular, we study the relic density as a function of the parameters of the model, and we find the  regions that are compatible with the dark matter constraint. Remarkably, this intermediate framework suggests a connection between dark matter and inflation, and  it allows, in contrast to WIMP dark matter, for dark matter particles of arbitrary mass, so even super-heavy dark matter becomes viable.

In the next section we explain in more detail the differences between the various theoretical frameworks that have been postulated to account for the dark matter. Then, we describe the intermediate scenario, paying particular attention to the assumptions behind it and to the computation of the relic density. In section \ref{sec:model}, we consider an explicit realization of this framework based on the singlet scalar model of dark matter.  The relic density is computed and  the new viable region corresponding to this dark matter scenario is obtained. It is then shown that this intermediate framework allows for super-heavy dark matter and that it suggests a connection between dark matter and inflation. Finally, some important implications of this intermediate framework are briefly discussed.

\section{WIMP, FIMP and EWIP dark matter}
\label{sec:wimp}
WIMPs are the most common candidates proposed to account for the dark matter of the Universe. The reason they are so popular is twofold. On the one hand, massive particles ($M\sim 0.1-1\tev$) with weak-strength interactions frequently appear in well-motivated extensions of the Standard Model. On the other hand, such particles naturally have a relic density not far from the observed dark matter density --the so-called WIMP miracle. Neutralinos in the MSSM are the most prominent example of this class. Due to their interactions, WIMPs easily reach thermal equilibrium in the early Universe and their relic density is the result of a freeze-out. The WIMP distribution simply follows the equilibrium distribution until the freeze-out temperature ($T\sim M/25$) is reached; from then on  the WIMP abundance remains essentially unchanged until today. The Boltzmann equation that determines the evolution of the WIMP density, $Y=n/s$, is
\begin{equation}
 \frac{dY}{dT}=\sqrt{\frac{\pi g_*(T)}{45}}M_p\sv (Y_{eq}(T)^2-Y(T)^2)
\label{eq:boltzmann}
\end{equation}
with the initial condition that WIMPs are in equilibrium at high temperatures: $Y(T_i\gtrsim M)=Y_{eq}(T_i\gtrsim M)$. An approximate analytical solution of this equation yields \cite{Griest:2000kj}
\begin{equation}
 Y(T_0)\approx 10^{-8}\frac{\gev}{M_{WIMP}} \frac{3\times 10^{-27}\svunits}{\sv}\,.
\end{equation}
From it, the present relic density of dark matter is obtained as
\begin{equation}
 \oh=0.1\frac{3\times 10^{-26}\svunits}{\sv}\,,
\end{equation}
which shows that the relic density is inversely proportional to the dark matter annihilation cross section and that the WMAP measurement \cite{Komatsu:2010fb} implies $\sv\approx 3\times 10^{-26}\svunits$. An important feature of WIMPs is that, thanks to their interaction strength and mass, they can be probed in different ways. They   can be produced and observed at colliders such as the LHC, or  be detected as they scatter of nuclei in direct detection experiments, or in indirect detection experiments through  their annihilation products --mainly gamma rays, antimatter, and neutrinos. Most of the dark matter experiments running today, in fact, were designed to detect  WIMP dark matter. As it is clear from this discussion, the WIMP framework is indeed a simple,  predictive, and verifiable  scenario for dark matter. 

FIMP dark matter is an alternative scenario in which the interactions of the dark matter particle, even though renormalizable, are of much weaker strength. As a  result, FIMPs do not reach equilibrium in the early Universe. In this case, dark matter particles are slowly produced by scatterings in the thermal plasma and they are never abundant enough to annihilate with each other. This production ceases, the dark matter abundance \emph{freezes-in}, when $T\lesssim M_{FIMP}$ and the particles in the plasma no longer have enough energy to produce dark matter \cite{Hall:2009bx}. The simplest realization of this framework is that of the singlet scalar model \cite{Yaguna:2011qn}. The equation that determines the evolution of the dark matter abundance in this case is      
\begin{equation}
\frac{dY}{dT}=\sqrt{\frac{\pi g_*(T)}{45}}M_p\langle\sigma v\rangle Y_{eq}(T)^2
\label{eq:boltzmann2}
\end{equation}
with the initial condition $Y(T_i\gg M)\sim 0$. Because their production is dominated by the region $T\sim M_{FIMP}$, the precise value of $T_i$ is not relevant for the calculation. For FIMPs, the dark matter relic density is proportional to $\sv$ and its observed value can be obtained for masses in the $\gev$ to $\tev$ range and couplings of order $10^{-11}-10^{-12}$ \cite{Yaguna:2011qn}. Such small couplings preclude the observation of FIMP dark matter in direct and indirect detection experiments. If, within the next decade, such experiments do not provide evidence of dark matter, the WIMP paradigm will have to be abandoned \cite{Bertone:2010at} and FIMP dark matter may become the most suitable scenario to account for the absence of such signals.  Conversely, if such evidence is found, FIMPs can be ruled out as the right explanation for dark matter. Notice that, in any case,   regarding dark matter the FIMP framework is as simple and predictive as the WIMP one. They both assume the Standard Cosmological Model and  their only difference is the typical interaction strength of the dark matter particle.

A third theoretical framework that can explain  the dark matter is that of EWIPs. Their main difference with respect to FIMPs is that in this case the interactions of the dark matter particle are non-renormalizable. The gravitino and the axino are the two typical examples of EWIPs. In this case, the relic density is also determined by equation (\ref{eq:boltzmann2}) and with the same initial condition.  Now, however, due to the non-renormalizable interaction, the production is dominated by the high temperature regime (rather than the low one) with the result that the relic density depends on the reheating temperature of the Universe, $\trh$. The gravitino relic density, for instance, is given by \cite{Bolz:2000fu}
\begin{equation}
 \oh = 0.27\left(\frac{\trh}{10^{10}\gev}\right)\left(\frac{100\gev}{\mgrav}\right)\left(\frac{\mglu(\mu)}{1\tev}\right)^2\,,
\end{equation}
where $\mgrav$ and $\mglu$ are respectively the gravitino and the gluino masses. This scenario is then slightly less predictive than the two previous ones, as it does not allow to compute the relic density in terms of the masses and couplings of the model. It opens the possibility, nevertheless, of determining the reheating temperature of the Universe via measurement at colliders \cite{Choi:2007rh}.

Whether dark matter consists of WIMPs, FIMPs, EWIPs or something else is at the end an experimental issue, and it is an issue that is being  addressed right now by a number of direct and indirect detection experiments, and by collider searches at the LHC.  The main point of this paper is to examine another theoretical framework that may account for the dark matter. Since this scenario has certain similarities with the three we have already discussed, we have dubbed it the intermediate framework between WIMP, FIMP and EWIP dark matter, or just the intermediate framework for short. 

\section{The intermediate framework}
\label{sec:int}
The intermediate dark matter framework shares some features with WIMP, FIMP and EWIP dark matter. The basic idea is to have a massive and weakly interacting particle --\'a la WIMP-- that does not reach thermal equilibrium in the early Universe --\'a la FIMP. To make these two conditions compatible with each other,  a reheating temperature smaller than the dark matter mass is assumed. In consequence, the dark matter relic density turns out to depend on the reheating temperature of the Universe --\'a la EWIP. 

The Boltzmann equation that determines the abundance of dark matter is the same as in the FIMP scenario:    
\begin{equation}
\frac{dY}{dT}=\sqrt{\frac{\pi g_*(T)}{45}}M_p\langle\sigma v\rangle Y_{eq}(T)^2\,,
\label{eq:boltz}
\end{equation}
but the initial condition is different. Now we require that $Y(\trh)=0$ with $\trh\ll M$, being $M$ the mass of the dark matter particle\footnote{Also in EWIP scenarios is assumed that $Y(\trh)=0$.}. It must be emphasized that this initial condition is one of the defining assumptions of the intermediate scenario we are examining. If this condition is not satisfied in a particular model, then such a model is simply not realized within the intermediate framework. It may indeed happen that $\trh\gg M$ as in the standard scenario, or that dark matter particles are abundantly produced during the reheating process so that $Y(\trh)=0$ is not a good approximation --see e.g. \cite{Gelmini:2006pq}. The validity of such initial condition, in fact, is ultimately related to inflationary models. Throughout this work, we simply assume that this condition is satisfied. 

Solving the above equation we can determine the dark matter abundance at low temperatures, $Y(T_0)$, and from it the relic density is found via
\begin{equation}
\oh=2.742\times 10^8 \frac{M}{\gev} Y(T_0), 
\label{eq:rd}
\end{equation}
where $M$ is the mass of the dark matter particle.
Since the temperature in equation (\ref{eq:boltz}) is always smaller than the mass of the dark matter particle, the equilibrium abundance is exponentially suppressed, $Y_{eq}\propto e^{-M/T}$. Thus, the relic density will depend exponentially on the reheating temperature. In addition, $\oh\propto \sv$ just as it happens for FIMPs. Hence, to compute the dark matter  relic density in this scenario we just need to know $M$, $\trh$ and the function $\sv$.  The intermediate framework can be realized within different particle physics models of dark matter. Each model will give a specific value for $\sv$ and a preferred range for the dark matter mass. Imposing the relic density constraint will then provide a relation between the reheating temperature of the Universe and the other parameters of the model.

To our knowledge, this intermediate framework of dark matter has not been analyzed in detail before. Non-standard scenarios for dark matter, including models with low reheating temperatures \cite{Gelmini:2006pq,Drees:2006vh} or with non-thermal production \cite{Arcadi:2011ev}, have certainly been considered in previous works, but, as we will see, they differ in significant ways from the intermediate scenario we are examining.  

In the next section we study a particular realization of the intermediate framework based on the singlet scalar model of dark matter. This model has the advantage that the function $\sv$ depends on only one additional coupling, so the relic density is entirely determined by three parameters\footnote{In other dark matter models the number of parameters is usually larger so the analysis becomes less transparent.}.

\section{A specific model}
\label{sec:model}
\begin{figure}[tb]
\mbox{
  \subfigure{\includegraphics[scale=0.21]{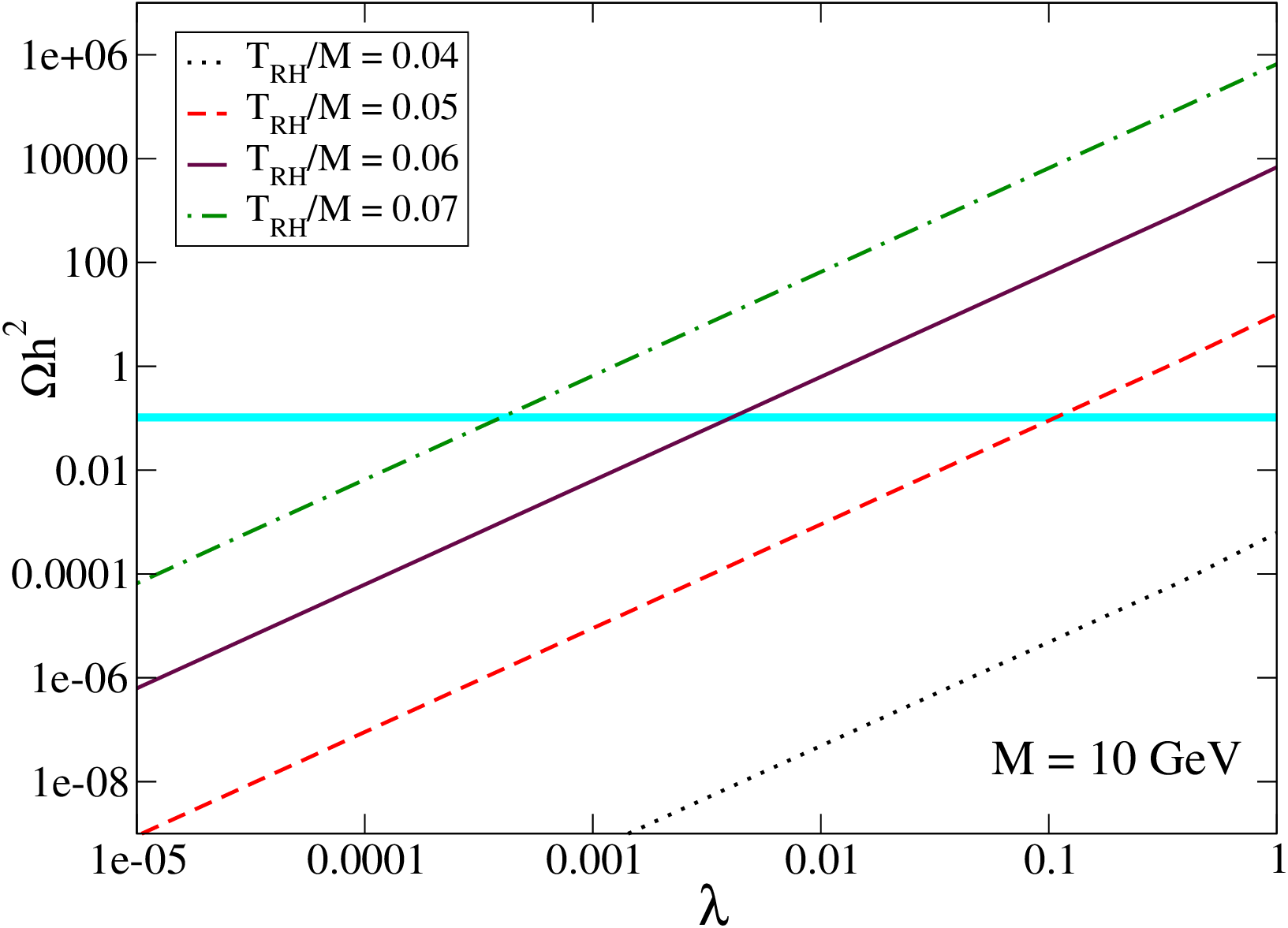}}
   \quad
  \subfigure{\includegraphics[scale=0.21]{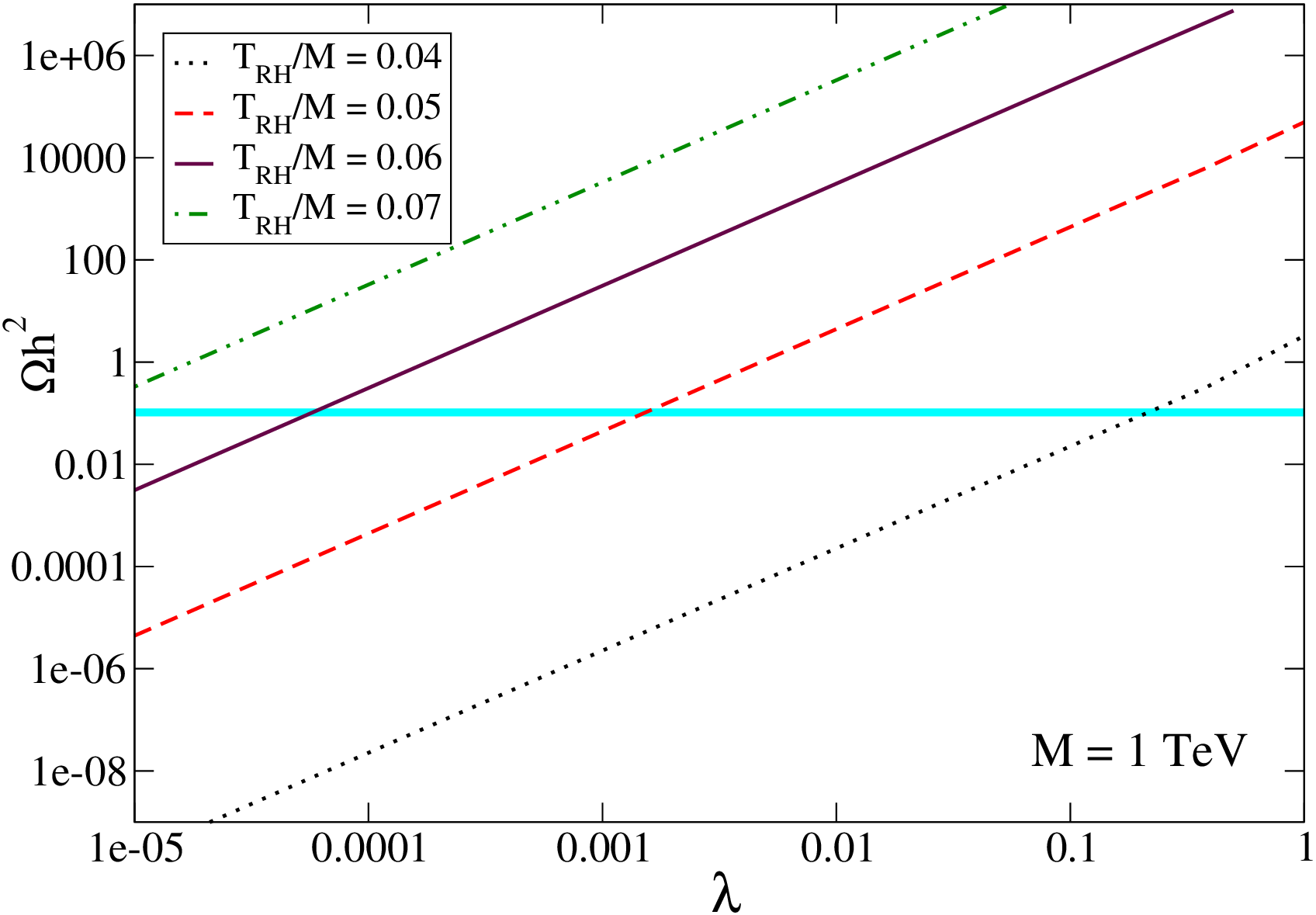}}
}
\caption{The relic density as a function of $\lambda$ for $M= 10 \gev$ (left) and $M= 1 \tev$. The lines corresponds to different values of $\trh/M$: $0.04$ (dotted line), $0.05$ (dashed line), $0.06$ (solid line), $0.07$ (dash-dotted line). The WMAP range is illustrated as a horizontal band.} \label{fig:omlam}
\end{figure}

The singlet scalar model \cite{McDonald:1993ex,Burgess:2000yq}  is one of the minimal extensions of the Standard Model that can explain the dark matter. It contains an additional field, $S$, that is singlet under the SM gauge group and odd under a new $Z_2$ symmetry that guarantees its stability. The Lagrangian that describes this model is
\begin{equation}
\mathcal{L}= \mathcal{L}_{SM}+\frac 12 \partial_\mu S\partial^\mu S-\frac{m_0^2}{2}S^2-\frac{\lambda_S}{4}S^4-\lambda S^2 H^\dagger H\,,
\label{eq:la}
\end{equation}
where $\mathcal{L}_{SM}$ denotes the Standard Model Lagrangian and $H$ is the higgs doublet. The above is the most general renormalizable Lagrangian that is compatible with  the assumed symmetries and field content of the model. The singlet scalar model introduces, therefore, only two relevant parameters\footnote{$\lambda_S$ only affects the self-interactions of the dark matter particle.}: the singlet mass ($M=\sqrt{m_0^2+\lambda v^2}$) and the coupling to the higgs boson, $\lambda$. In addition, the higgs mass, a SM parameter, also affects the phenomenology of the model. Given the narrow range over which $m_h$ is allowed to vary \cite{CMS-PAS-HIG-11-023}, we will in the following take $m_h=120\gev$. Our results do not strongly depend on this choice.

In the early Universe, singlets are pair-produced as the particles in the thermal plasma scatter off each other. The dominant production processes are  $s$-channel higgs boson mediated diagrams originating in   a variety of initial states: $f\bar f$, $W^+W^-$, $Z^0Z^0$, and $hh$. Likewise, they can also be produced from the initial state $hh$ either directly or through singlet exchange. As a general rule, it is the initial state $W^+W^-$ that tends to dominate the total production rate of dark matter in this model.

The main advantages of the singlet model with respect to other models of dark matter are its simplicity, which allows one to make concrete predictions about dark matter observables, and its versatility, which allows us to use it as a toy model for many different dark matter studies. In previous works, it was shown that the singlet model can explain the dark matter either in the WIMP regime \cite{Yaguna:2008hd,Goudelis:2009zz} or in the FIMP regime \cite{Yaguna:2011qn}. Here, we will show that it can also account for the dark matter in the intermediate framework.

The Boltzmann equation that determines the dark matter abundance, equation (\ref{eq:boltz}), can be solved either analytically or numerically. The analytic solution relies on the velocity expansion of the annihilation rate --see e.g. \cite{Griest:2000kj}-- which, as is well-known, breaks down close to resonances and thresholds. Since in the singlet model the higgs resonance and the $W^\pm$ (and $Z^0$) threshold play a very important role in the evaluation of $\sv$, the analytic solution becomes unreliable and one has to resort to  numerical methods. In our calculations we have used micrOMEGAs \cite{Belanger:2010gh} to compute $\sv$, as it automatically takes into account resonance and threshold effects. With it,  equation (\ref{eq:boltz}) can be easily solved numerically to obtain $Y(T_0)$ and $\oh$.  

\begin{figure}[tb]
\includegraphics[scale=0.4]{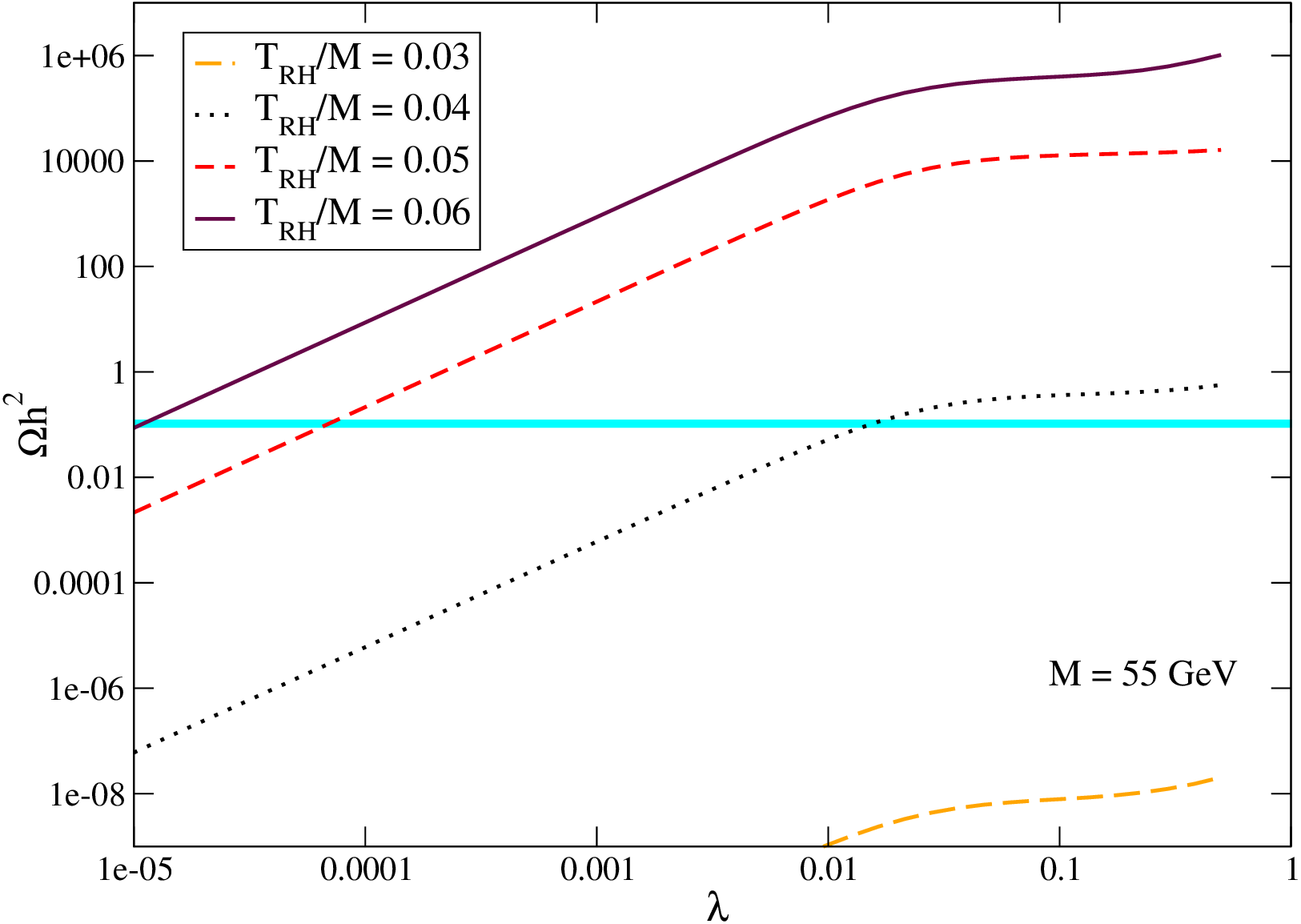} 
\caption{The relic density as a function of $\lambda$ for a dark matter mass ($M= 55 \gev$) close to the higgs resonance ($m_h/2=60\gev$).  The lines corresponds to different values of $\trh/M$: $0.03$ (dotted line), $0.04$ (dashed line), $0.05$ (solid line), $0.06$ (dash-dotted line). The WMAP range is illustrated as a horizontal band.} \label{fig:omlam55}
\end{figure}

In this scenario, the singlet relic density depends on three parameters: $\lambda$, $M$ and $\trh$. It is convenient however to use $\trh/M$ rather than $\trh$ as the free parameter. Figure \ref{fig:omlam} shows, for $M=10\gev$ (left) and $M=1\tev$ (right), the predicted relic density as a function of $\lambda$ for different values of $\trh/M$. As expected, $\oh$ increases quadratically with $\lambda$. It also increases, though significantly faster, with the ratio $\trh/M$. In fact, from $\trh/M=0.04$ to $\trh/M=0.07$ the relic density changes by about $9$ orders of magnitude, a consequence of the exponential suppression in the equilibrium density of dark matter. The behavior of $\oh$ is analogous for $M=10\gev$ and $M=1\tev$, the only difference being a value about four orders of magnitude larger for  the latter. Notice, in particular, that if we want to have dark matter with weak-strength interactions ($\lambda\sim 0.1$), $\trh/M$ should be about $0.05$ for $M=10\gev$ or $0.04$ for $M=1\tev$. It is clear from the figure, however, that the correct density of dark matter can also be obtained for much smaller couplings with only a mild increase in the ratio $\trh/M$.  

\begin{figure}[tb]
\includegraphics[scale=0.4]{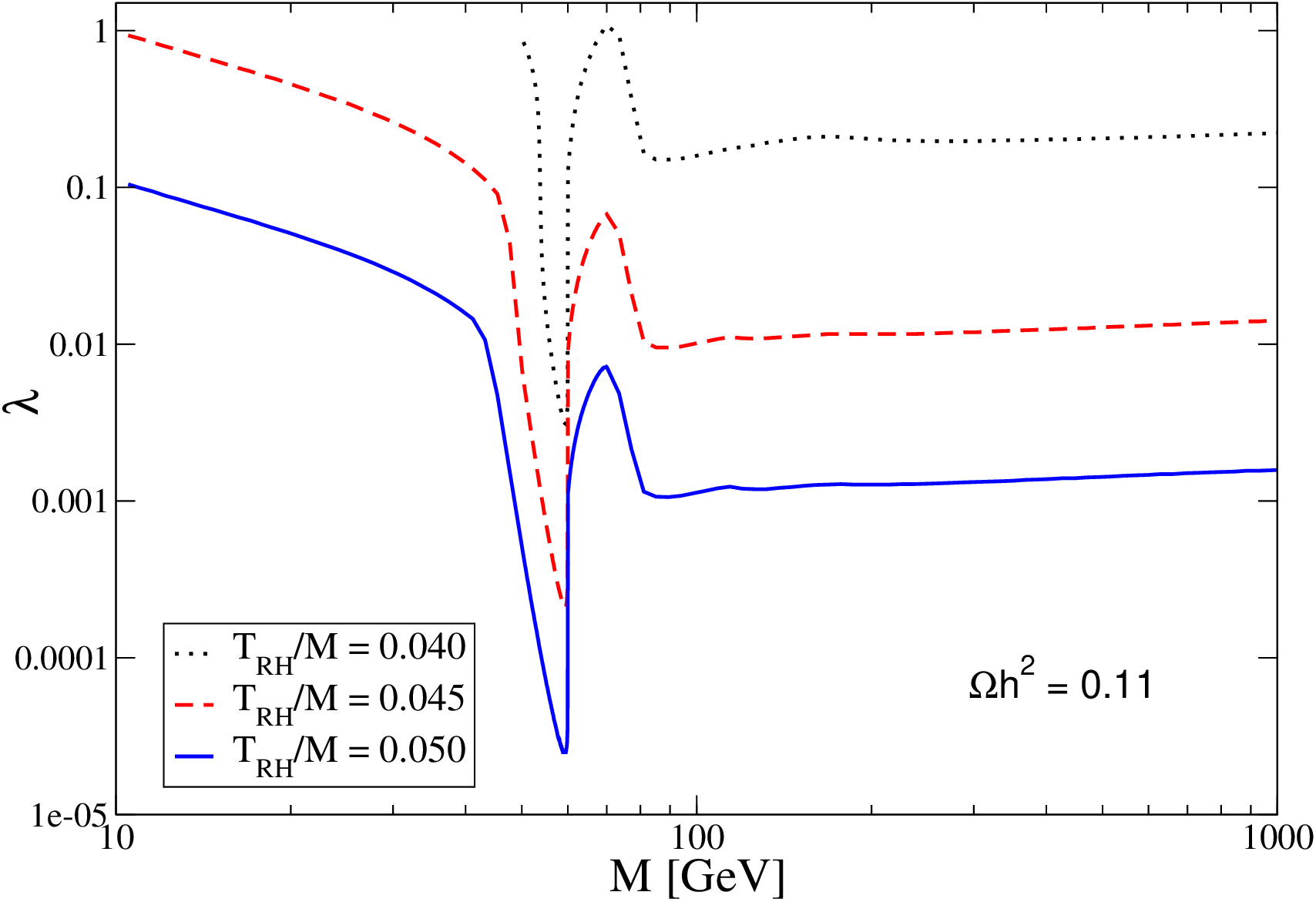} 
\caption{The viable parameter space of the singlet scalar model in the plane ($M$, $\lambda$) for different values of $\trh/M$. From top to bottom the lines correspond to $\trh/M= 0.040$ (dotted line), $0.045$ (dashed line), and $0.050$ (solid line). This viable region corresponds to the intermediate regime between WIMP, FIMP and EWIP dark matter examined in this paper.\label{fig:pspace}}
\end{figure}

If the dark matter mass is close to the higgs resonance, $\sv$ is enhanced and the production of dark matter particles is increased, leading to a different behavior of the relic density --see figure \ref{fig:omlam55}. In this figure $M=55\gev$, so there is only a $5\gev$ splitting between the dark matter mass and the higgs resonance ($m_h/2=60\gev$). The effect of the resonance is twofold: it yields a larger value of $\oh$ over the entire range of $\lambda$ and $\trh/M$, and it gives rise to a deviation from a straight line for $\lambda\gtrsim 0.01$. In that region, the production cross section is large enough to allow the total conversion of radiation into dark matter; essentially all particles in the thermal plasma with energies large enough to annihilate into a pair of dark matter particles do so, giving rise to an asymptotic value of the relic density. Indeed, the relic density hardly increases for $\lambda>0.01$.

 \begin{figure}[tb]
\includegraphics[scale=0.4]{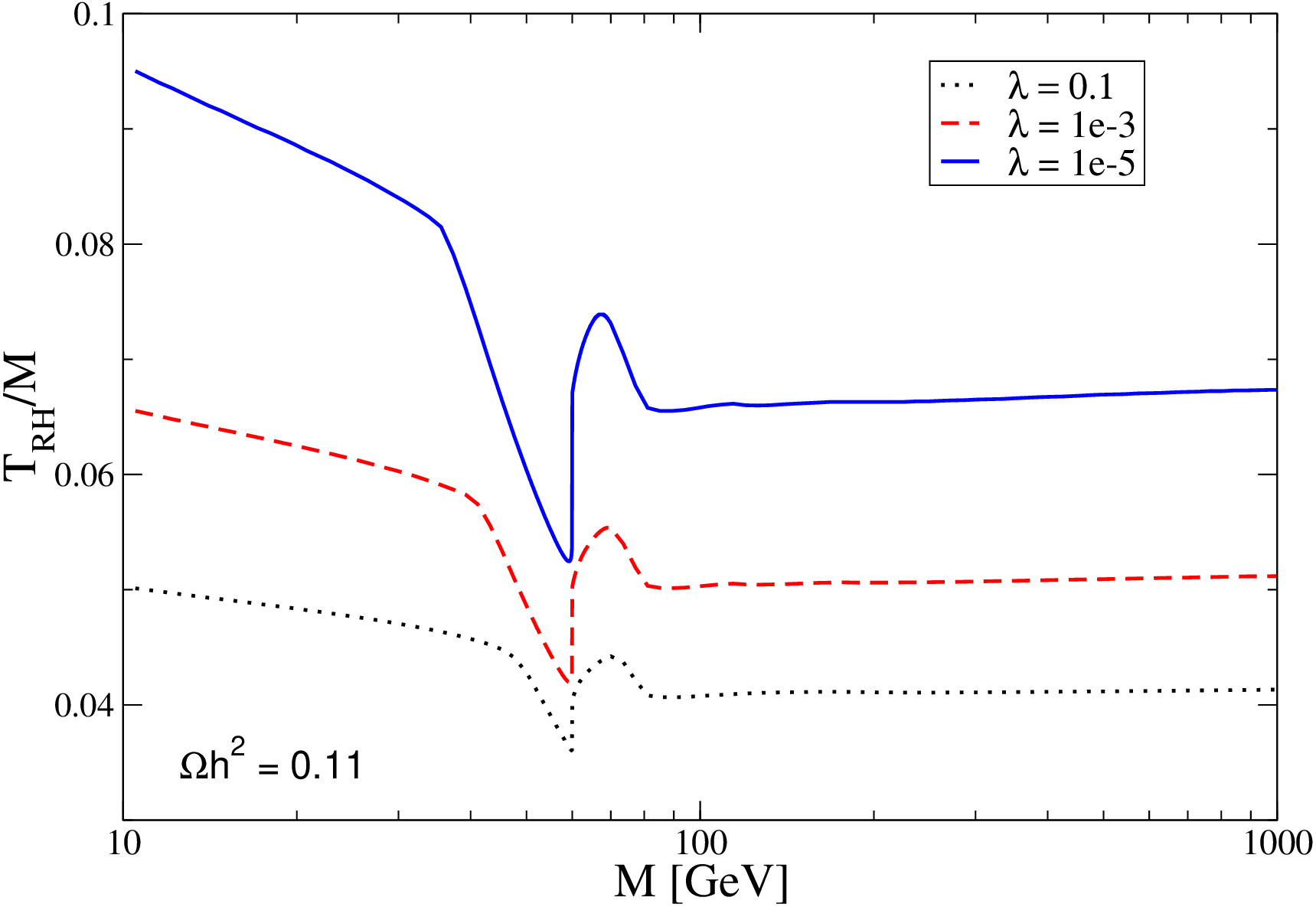} 
\caption{The viable parameter space of the singlet scalar model in the plane ($M$, $\trh/M$) for different values of $\lambda$. From top to bottom the lines correspond to $\lambda= 10^{-5}$ (solid line), $10^{-3}$ (dashed line), and $10^{-1}$ (dotted line).This viable region corresponds to the intermediate regime between WIMP, FIMP and EWIP dark matter examined in this paper. \label{fig:pspace2}}
\end{figure}

 If we now impose the relic density constraint, $\oh=\oh_{WMAP}$, one of the three  parameters that determines the relic density can be eliminated, and the viable parameter space of the model is obtained. Figure \ref{fig:pspace} shows the viable parameter space in the plane ($M$, $\lambda$) for different values of $\trh/M$: 0.040,0.045, 0.050. In it, we have restricted  $\lambda$ to be smaller than $1$ so that perturbativity is guaranteed. The curves all have the same behavior: for $M\lesssim 50\gev$ $\lambda$ decreases with the dark matter mass whereas for $M\gtrsim 100\gev$ $\lambda$ remains almost constant --it  actually increases slightly. In the region $50\gev<M<100\gev$ the behavior is more complicated due to the presence of the higgs resonance and the $W^\pm$ (and $Z^0$) threshold.  The required value of $\lambda$ changes over several orders of magnitude (decreasing and then increasing in the higss resonance and finally decreasing once more at the $W^\pm$  threshold) within that small mass range. As expected, the larger the value of $\trh/M$ the smaller the viable $\lambda$. For the values of $M$ and $\trh/M$ we consider, the correct dark matter density is typically obtained for $\lambda$ between $1$ and $10^{-3}$.

One can also study the viable parameter space of this model in the plane ($M$, $\trh/M$) for different values of $\lambda$ --see figure \ref{fig:pspace2}. In it we consider $\lambda=0.1, 10^{-3}, 10^{-5}$. Again, three different mass ranges can easily be distinguished: $M\lesssim 50\gev$, $M\gtrsim 100\gev$, and the region in-between. For $M\lesssim 50\gev$, $\trh/M$ decreases with $M$ whereas it remains almost constant for $M\gtrsim 100\gev$. As observed in the figure, for $50\gev<M<100 \gev$ the value of $\trh/M$ varies apreciably. Notice, however, that the variation in $\trh/M$ is much smaller than that in $\lambda$: for $\lambda$ in the range  ($10^{-5},0.1$) $\trh/M$ varies only from $0.04$ to $0.1$. At higher masses ($M>100\gev$), the range of variation is even smaller --from $0.04$ to $0.07$.

In conclusion,  the intermediate framework allows to account for the observed dark matter density within a restricted range of $\trh/M$ but over  a wide range of couplings and  masses. Part of that region overlaps with that characteristic of the WIMP scenario: masses in the $\gev$-$\tev$ range and couplings of order $10^{-2}-1$. Hence, all WIMP models considered in the literature, including neutralino dark matter, can also be realized within the intermediate scenario. That is, for any given SUSY model it is possible to find a value of $\trh/M$ such that the relic density is in agreement with the observations. Such values would then define the new viable regions for neutralino dark matter in this framework. More interesting, however, is the fact that the intermediate framework opens up new possibilities not found within the WIMP scenario, as we show next. 

\section{Super-heavy dark matter}
\label{sec:heavy}
In the usual WIMP scenario, the dark matter mass cannot be arbitrarily large. A model-independent upper bound, based on unitarity, in fact gives $M\lesssim 100 \tev$ \cite{Griest:1989wd}. And in specific models, such as the CMSSM, the bound is typically stronger. This upper limit on the mass of the dark matter particle, however, applies only to WIMP dark matter. It is natural, therefore, to ask whether one can have much heavier dark matter particles, that is super-heavy dark matter,  in the intermediate framework we are considering. We will see that that is indeed the case. The intermediate scenario can easily accommodate super-heavy dark matter.

\begin{figure}
\includegraphics[scale=0.4]{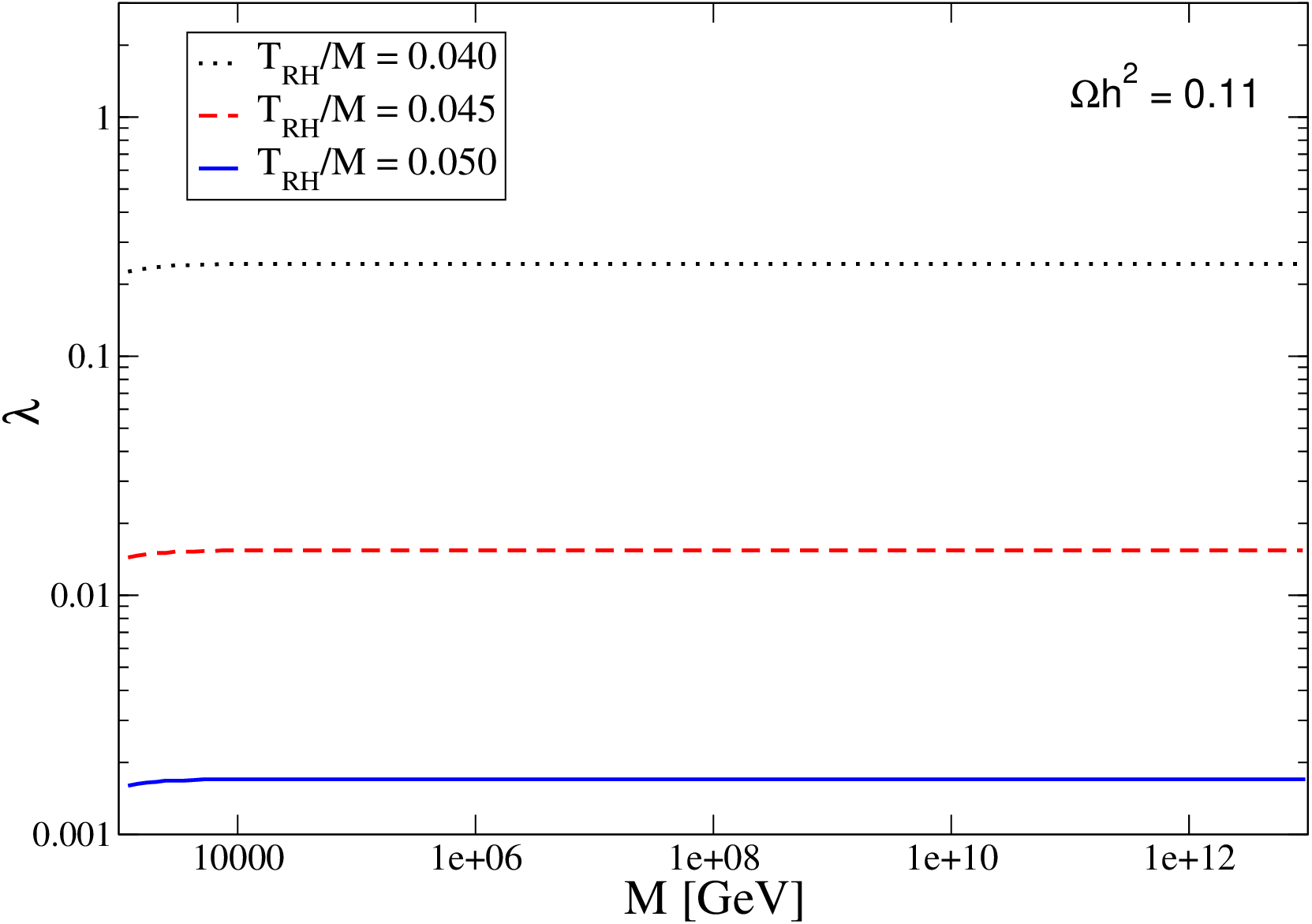}
\caption{The viable parameter space of the  model in the plane ($M$, $\lambda$) for the super heavy dark matter regime. The lines correspond, from top to bottom, to different values of $\trh/M$: $0.040$ (dotted line), $0.045$ (dashed line), $0.050$ (solid line).\label{fig:pspaceh}} 
\end{figure}
\begin{figure}
\includegraphics[scale=0.4]{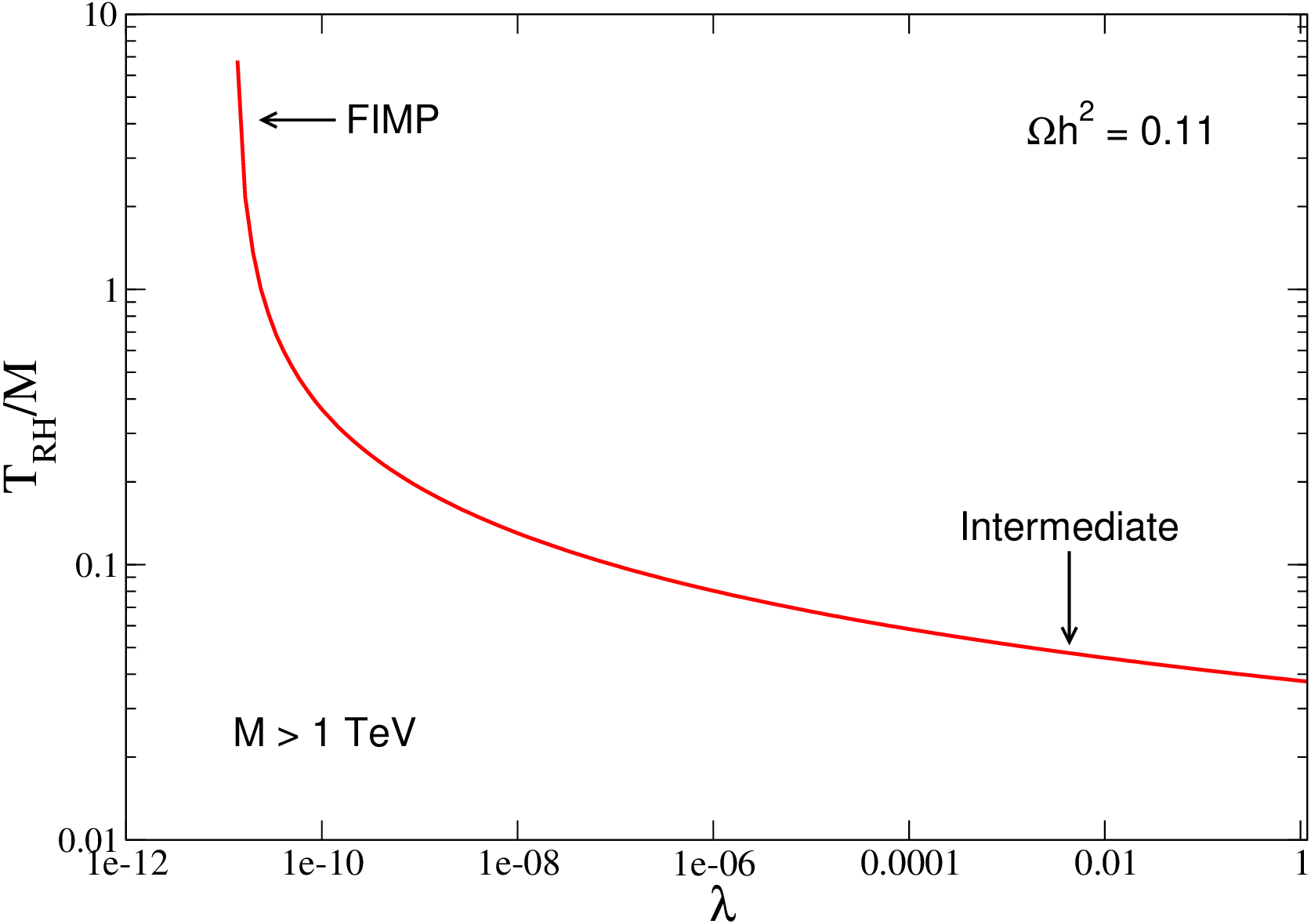} 
\caption{The viable parameter space of the  model in the plane ($\lambda$, $\trh/M$) for the super heavy dark matter regime. This viable region does not depend on $M$ for $M\gtrsim 1\tev$. Notice that the range of $\lambda$ and $\trh/M$  considered is much larger in this figure than in all the previous ones.\label{fig:pspaceh3}} 
\end{figure}

Figure \ref{fig:pspaceh} shows, for the super-heavy dark matter regime, the viable parameter space of the model in the plane ($M$, $\lambda$) for three different values of $\trh/M$. Notice that in the figure the dark matter mass varies between $1\tev$ and $10^{13}\gev$. The required value of $\lambda$ lies approximately between $10^{-3}$ and $1$ (for the examined values of $\trh/M$), and it is observed  not to depend on the dark matter mass --it is entirely determined by $\trh/M$. It makes sense, henceforth, to illustrate the viable parameter space rather in the plane ($\lambda$, $\trh/M$) for \emph{any} value of $M$, as shown in figure \ref{fig:pspaceh3}. In it, we have considered a much wider range for $\lambda$ and $\trh/M$ so that the transition from the intermediate framework analyzed in this paper to the FIMP regime ($\lambda\sim 10^{-11}$ , $\trh/M$ large) can be observed. As expected, at high values of $\trh/M$ the required value of $\lambda$ becomes independent of $\trh/M$ and equal to that obtained for  FIMP dark matter. Notice that for $\lambda\gtrsim 10^{-8}$ the variation in $\trh/M$ is quite small, in agreement with our previous results. The high mass regime of the intermediate framework thus predicts a well-defined relation between $\lambda$ and $\trh/M$. If $\trh/M$ can be obtained, say from a given inflationary model, then one could use the above figure to determine the coupling $\lambda$ and consequently the interaction between the dark matter and the Standard Model particles. Additional input would be needed, however, to obtain some  information on the dark matter mass.

An important lesson of this super-heavy regime is that the solution to the dark matter puzzle is not necessarily associated with new physics at the $\tev$ scale. Contrary to current expectations, dark matter might be the result of physics at scales well-above the reach of present and planned experiments.

\section{Discussion}
\label{sec:disc}
The intermediate framework does not make any specific prediction regarding the detectability of dark matter. 
In general, the detection prospects of the dark matter particle  depend on its mass and its interactions with other SM particles, that is on $M$ and $\lambda$ for the singlet scalar model we have studied. As we have seen, the intermediate framework allows these two parameters to vary over several orders of magnitude, so the detectability of singlet dark matter will depend on the specific values that are realized in nature. If, for instance, $M\sim 0.1-1\tev$ and $\lambda\gtrsim 10^{-2}$, the detection prospects would be similar to those expected for WIMPs. Dark matter could be observed in direct and indirect detection experiments and be produced at high energy colliders such as the LHC. In this optimistic case, one could in principle reconstruct $\sv$ from the data and use it to compute $\oh$ within the Standard Cosmological Model \cite{Baltz:2006fm}. If that value does not agree with the observed one, we would know that some new phenomena enters into the determination of the relic density and the intermediate framework would provide a plausible solution --certainly not the only one-- to that disagreement. But, if the couplings are smaller or the masses are larger the detection prospects of singlet scalar dark matter quickly fade away. More concrete predictions for the detection prospects of dark matter in direct and indirect detections experiments can be made in other models of dark matter where the range of variation of the relevant parameters is not as wide as in the singlet scalar model.

An interesting aspect of the intermediate framework is the connection that it entails between $\trh$ and $M$. We have found that this scenario can account for the observed dark matter density if the reheating temperature of the Universe and the mass of the dark matter particle are related by
\begin{equation}
\trh\sim 4-5\,\times 10^{-2} M
\end{equation}
for typical values of the coupling\footnote{This result was obtained for the singlet model but it can clearly be generalized to other dark matter models.}. That $\trh$ and $M$,  two apparently unrelated quantities, should be related to one another by  a number not much smaller than one is a very suggestive fact. It points toward a link between two of the most important open problems in fundamental physics: dark matter and inflation. A link that becomes more easily realized in the super-heavy dark matter regime of this framework. In fact, for $M$ in the $\gev$ to $\tev$ range the above condition would require a reheating temperature significantly below the electroweak scale --that is, a scenario with \emph{low} reheating temperature. But in the super-heavy regime, the reheating temperature can be as large as usually assumed in the Standard Cosmological Model, $\trh\sim 10^{8}-10^{12}\gev$. It is clear, therefore, that the intermediate framework examined in this paper does not require a low reheating temperature --although it allows it. This is an important feature from the point of view of inflationary models because  scenarios with low reheating temperatures are being constrained by CMB data \cite{Martin:2010kz}. Furthermore, this link to inflationary models might open new possibilities to test the intermediate framework via cosmological observations. In a future work we will investigate in more detail this intriguing connection between dark matter and inflation.

\section{Conclusions}
We have considered the intermediate framework for dark matter, a scenario that shares some features with WIMP, FIMP and EWIP dark matter. It consists of a weakly interacting particle (WIMP-like) that is unable to reach thermal equilibrium with the thermal plasma (FIMP-like) because the reheating temperature of the Universe is smaller than the dark matter mass. Consequently, the dark matter relic density depends also on $\trh$ (EWIP-like). An explicit realization of this scenario, based on the singlet scalar model of dark matter, was analyzed in detail. Specifically, the region of the parameter space that is compatible with the dark matter constraint within this scenario was determined. A salient feature of this framework is that, in contrast to WIMP dark matter, it allows for dark matter particles of arbitrary mass, so even super-heavy dark matter is viable. Moreover, since the scenario implies that $\trh$ and $M$ are connected to one another, it  suggests a thought-provoking link between dark matter and inflation.  

\section*{Acknowledgments}
This work is supported by the ``Helmholtz Alliance for Astroparticle Phyics HAP''
 funded by the Initiative and Networking Fund of the Helmholtz Association.

\bibliographystyle{hunsrt}
\bibliography{darkmatter}

\end{document}